\documentclass[twocolumn]{aastex631}

\usepackage{graphicx}	
\usepackage{amsmath}	
\usepackage{amssymb}	
\usepackage{bm}		
\usepackage{subfigure}
\usepackage{hyperref}
\usepackage{booktabs}
\usepackage{tablefootnote}

\newcommand{\kms}{\,km\,s$^{-1}$} 
\newcommand{\tco}{$^{13}$CO(3-2)}
\newcommand{\cetno}{C$^{18}$O(3-2)}
\newcommand{\co}{CO(3-2)}
\newcommand{\comment}[1]{}

\shorttitle{Sh2-138 Hub-Filament System}
\shortauthors{K K Mallick et al.}

\graphicspath{{./}{figures/}}

\begin{document}

\title{Structure and Kinematics of Sh2-138 - A distant hub-filament system in the outer Galactic plane}


\author[0000-0002-3873-6449]{Kshitiz K. Mallick}
\affiliation{Aryabhatta Research Institute of Observational Sciences (ARIES), \\
             Manora Peak, Nainital, 263002, India}

\author[0000-0001-6725-0483]{Lokesh K. Dewangan}
\affiliation{Physical Research Laboratory, Navrangpura, Ahmedabad 380009, India}

\author[0000-0001-9312-3816]{Devendra K. Ojha}
\affiliation{Department of Astronomy and Astrophysics, \\
             Tata Institute of Fundamental Research, Homi Bhabha Road, \\
             Mumbai 400005, India}

\author[0000-0003-0295-6586]{Tapas Baug}
\affiliation{S.N. Bose National Centre for Basic Sciences, Block JD, \\
             Sector III, Salt Lake, Kolkata 700106, West Bengal, India}

\author[0000-0003-2793-8229]{Igor I. Zinchenko}
\affiliation{Institute of Applied Physics of the Russian Academy of Sciences 46 Ul'yanov str., \\
             603950 Nizhny Novgorod, Russia}


\begin{abstract}
We present a molecular line study of the Sh2-138 (IRAS\,22308+5812)
``hub-filament'' system with an aim to investigate its structure and
kinematics.
Archival CO molecular line data from the Canadian Galactic Plane Survey
(CO(J=1--0)) for the wider region ($\sim$\,50\arcmin$\times$50\arcmin) and
the James Clerk Maxwell Telescope (\co, \tco, and \cetno) for the central
portion ($\sim$\,5\arcmin$\times$5\arcmin) have been utilised.
Analysis of the CO(1-0) spectra for the extended region in conjunction
with the hub and filament identification using column density map and
the \emph{getsf} tool, respectively, reveals a complex structure with
the spectral extraction for the central position displaying multiple
velocity components.
Based on the \emph{Herschel} 70\,\micron\, warm dust emission,
one of the filaments in the extended region was inferred to be
associated with active star formation, and is host to a Bolocam
1.1\,mm clump of $\sim$\,1606\,M$_\sun$.
Integrated intensity map of \tco\, emission, constructed from
clumps detected at above 5$\sigma$ in position-position-velocity space,
reveals three filamentary structures (labelled W-f, SW-f, and SE-f) in
the central portion.
Velocity gradients observed in \tco\, position-velocity slices point to
longitudinal gas flow along the filaments into the central region.
Filaments W-f, SW-f, and SE-f were calculated to have observed line masses
of $\sim$\,32, 33.5, and 50\,M$_{\sun}$\,pc$^{-1}$, respectively.
The cloud was found to be dominated by supersonic and non-thermal motions,
with high Mach numbers ($\gtrsim$\,3) and low thermal to non-thermal pressure ratio
($\sim$\,0.01--0.1).
\end{abstract}

\keywords{Interstellar medium (847), Interstellar filaments (842), Interstellar molecules (849), H II regions (694), Millimeter astronomy (1061), Star formation (1569), Massive stars (732)}


\section{Introduction}
\label{section_introduction}

Understanding the evolutionary sequence of high-mass stars ($\geq$8 M$_{\sun}$)
is an area of ongoing development in astronomy \citep{Zinnecker_Yorke_2007}.
Observations indicate that massive stars are generally found in
the center of embedded young stellar clusters in giant molecular clouds
\citep{Nanda_Clusters_AandA_2006,Beuther_PPV_2007,PortegisZwart_ARAA_YMCs_2010}.
The Lyman continuum radiation output of such stars ionizes the natal
cloud leading to the
formation of H\,II regions, which evolve from a compact to classical phase.
Observations in the past few decades have led to an empirical view that,
for high-mass stars, dense and massive cold cloud structures could be
analogous to prestellar cores in low-mass stars \citep{Motte_Review_2018}.
Infrared dark clouds -- seen in absorption against mid-infrared
emission (such as \emph{Spitzer}) -- are promising sites of such dense
cloud cores, with some resembling
the ``hub-filament'' model of \citet{MyersFilaments_09}
seen in emission in far-infrared observations by \emph{Herschel}.
According to \citet{Inutsuka_Miyama_1997}, filamentary features represent an
important step in the evolutionary life of a molecular cloud as it progresses
towards fragmenting into dense clumps.
The study by \citet{Nanda_2020} has even concluded that almost all massive
star formation occurs in the hubs and suggested a ``filaments to clusters''
paradigm.

It thus becomes an imperative to study young stellar clusters hosting high-mass
stars, with the presence of photodissociated region being their typical signature.
Sh2-138 (or G105.6270+00.3388;
\emph{l}\,$\sim$\,105.6270\arcdeg, \emph{b}\,$\sim$\,+0.3392\arcdeg),
associated with the IRAS source (IRAS\,22308+5812)
in Cepheus, is one such Galactic optical H\,II region.
The extent of the ionized region and the parameters derived (electron density,
emission measure, etc) from radio continuum
observations \citep{Fich_wrtS138_1993, MartinHernandez_wrtS138_2002} showed
it to be a classical H\,II region \citep{Kurtz_UCHIIRegions_2002}.
Early optical and near-infrared observations by \citet{Deharveng_S138_1999}
had found four O/B-type stars in a similar layout as the Orion Trapezium
cluster.
The multiwavelength study of \citet{Tapas_S138_2015} revealed further
compact radio clumps, as well as provided age and spectral type estimates
of the cluster of O/B stars.
Their analyses found an isolated cluster of young stellar objects -- with a
mean age of $\sim$\,1\,Myr -- centered on the location of the IRAS source
lying at the junction of filaments.
The massive star(s) seem to be driving molecular outflows in the region
\citep{Qin_wrtS138_2008}, and a possible (weak) water maser detection
\citep{Cesaroni_wrtS138_1988,Palagi_wrtS138_1993,Wouterloot_wrtS138_1993}
is likely an outcome of this \citep{Fish_masers_2007}, though more recent studies
\citep{Urquhart_wrtS138_2011} suggest non-detection.
It should however be noted that there is an absence of methanol masers
in the region \citep{Slysh_wrtS138_1999,Szymczak_wrtS138_2000}.
Early molecular observations of the region in
CO \citep{Dickinson_wrtS138_1974, Blitz_wrtS138_1982, Wouterloot_WB89_1989},
HCN \citep{Burov_wrtS138_1988},
HCO+ \citep{Zinchenko_wrtS138_1990,Yoo_wrtS138_2018},
NH$_3$ \citep{Harju_wrtS138_1993, Urquhart_wrtS138_2011},
CS \citep{Bronfman_wrtS138_1996}, and
other species' isotopologues \citep{Johansson_wrtS138_1994} were able
to detect the spectra of the molecular cloud centered in the
$\sim$-53 to -52\kms\, range.
More recent higher resolution CO observations \citep{Kerton_wrtS138_2003, Brunt_wrtS138_2003}
have found multiple velocity components associated with the molecular
cloud.
The multitude of molecular line studies have probed different physical conditions
in the cloud, but at specific locations (mostly on the coordinate of the IRAS source)
and have not looked at the spatial variation of the spectra (and thus the physical
conditions) in the wider region,
though this could be partly due to the low resolution of some of the studies.
The relatively comprehensive study of \citet{Tapas_S138_2015}
has focussed on optical and near-infrared study of stellar sources and the
ionized morphology. Also, their inference from \emph{Herschel} column density map
that this region lies at the junction of filaments makes it but natural to
explore the structure of the filamentary features in molecular line transitions.
The current paper aims to fulfill this void (partly, in various transitions of CO
isotopologues) for a better understanding of
the interplay of (massive) stellar cluster and hub-filament system (HFS)
in the Sh2-138 H\,II region.

\citet{Deharveng_S138_1999} use a distance of 5.0\,$\pm$\,1.0\,kpc for the
Sh2-138 H\,II region in their analysis based on the average distance to
nearby compact H\,II regions, though their kinematic distance calculation
from V$_{\textsc{lsr}}$(CO) suggests a value in the range 5.45-5.9\,kpc.
\citet{Anderson_wrtS138_2014} also calculated a value of $\sim$5.8\,kpc
based on the velocity from NH$_3$ spectrum.
According to \citet{Blitz_wrtS138_1982}, H\,II regions with similar velocities
($\sim$-53\kms) near Sh2-138, namely  Sh2-148 and Sh2-149, have distances of
$\sim$5.5 and 5.4\,kpc, respectively.
Similar distance estimates have been used by
\citet[$\sim$\,5.7\,kpc]{Wouterloot_WB89_1989};
\citet[$\sim$\,6.0\,kpc]{Johansson_wrtS138_1994} using a mean of kinematic distance and the estimate from
the size-linewidth relation;
\citet[$\sim$\,5.5\,kpc]{MartinHernandez_wrtS138_2002};
\citet[$\sim$\,5.7\,kpc]{Tapas_S138_2015}; and
\citet[$\sim$\,5.7\,kpc]{Zhang_wrtS138_2020}.
For the purpose of this paper, we thus adopt the distance of 5.7\,kpc of
\citet{Tapas_S138_2015}.

The organisation of the paper is as follows.
In section \ref{section_dataused}, we list the various datasets
used in this paper.
This is followed by
analysis and results in section \ref{section_AnalysisResults},
where the large scale view of the region is examined
(section \ref{section_largescaleview}) and
the kinematics of the molecular gas in the central region
is presented (section \ref{section_kinematics_JCMTobs})
A discussion of our results follows in section \ref{section_discussion}
and finally we conclude with a summary of major findings in
section \ref{section_summaryconclusions}.


\section{Data Used}
\label{section_dataused}

Table \ref{table_dataused} provides a summary of datasets used in this paper.
A brief description of the salient portions is being provided in the following
part of this section.

Archival spectral cubes for $^{12}$CO(1-0) (2.6\,mm)
emission produced by the Canadian Galactic Plane Survey
(CGPS) Consortium \citep{Taylor_2003_CGPS} were procured for the region.
The CGPS cubes are based on the Five College Radio Astronomy Observatory
(FCRAO) outer Galaxy survey \citep{Heyer_1998_FCRAO_CO} regridded to a
pixel scale of 18\arcsec\, and channel width of 0.824\kms.
The spectral cube data, given in radiation temperature scale (i.e. T$_\textsc{r}^*$),
has a spatial resolution of 100.44\arcsec.

Besides CGPS, archival JCMT (\emph{James Clerk Maxwell Telescope}) observations
of the Sh2-138 (G105.6270+00.3388) region were downloaded using the
CADC\footnote{\href{https://www.cadc-ccda.hia-iha.nrc-cnrc.gc.ca/en/search/}{https://www.cadc-ccda.hia-iha.nrc-cnrc.gc.ca/en/search/}}
data repository.
Calibrated spectral cubes for the three molecular lines --
\co (rest frequency = 345.79599\,GHz; Proposal ID : M07AU08; Int. time : 17.983\,s),
\tco (rest frequency = 330.587960\,GHz; Proposal ID : M08BU18; Int. time : 48.896\,s),
and
\cetno (rest frequency = 329.330545\,GHz; Proposal ID : M08BU18; Int. time : 48.808\,s)
-- observed using the HARP/ACSIS
\citep[Heterodyne Array Receiver Programme/Auto-Correlation Spectral Imaging System;][]{Buckle_HARP_ACSIS_2009}
spectral imaging system were retrieved.
The \emph{J=3--2} transition traces gas at a higher critical density
\citep[$\sim$\,10$^{4-5}$\,cm$^{-3}$]{Buckle_JLSGould_OrionB_2010} than the
CGPS CO(J=1--0) data whose critical density is of the order of
$\sim$\,10$^{3}$\,cm$^{-3}$ \citep{Bolatto_ARAA_COtoH2_2013, Shirley_CriticalDen_2015}.
The temperature scale used for the pixel brightness units is T$_\textsc{a}^*$ (antenna
temperature) for all the three spectral cubes.
Basic processing for the purpose of our analysis involved conversion of
spectral axis units from frequency to velocity scale, and the coordinate system
to Galactic from FK5.
Both the \tco\, and \cetno\, JCMT cubes had a channel width of
$\sim$\,0.05\kms, with the mean rms noise being $\sim$\,0.96$\pm$0.45\,K
and 1.3$\pm$0.7\,K, respectively.
While we use these cubes for spectra analysis due to their high velocity
resolution; for the detection of spatial structures, we averaged and
rebinned both these cubes along the spectral axis to 0.5\kms\, channels.
The resultant rebinned cubes had a reduced mean rms noise
of 0.30$\pm$0.14\,K and 0.41$\pm$0.23\,K
for \tco\, and \cetno, respectively.
The above tasks were implemented using the
\textsc{starlink kappa} \citep{Currie_starlink_2014} package
commands such as ``wcsattrib'' and ``sqorst''.
The \co\, cube -- which had a native channel width of $\sim$0.42\kms\, and
a mean rms noise of 0.53$\pm$0.2\,K -- was mainly used to examine the
morphology of the region.
Each of the images in the cubes has a pixel scale of $\sim$7.3\arcsec,
and a beamsize of $\sim$14\arcsec\, \citep{Buckle_HARP_ACSIS_2009,
Buckle_JLSGould_OrionB_2010, Davis_JLSGould_Taurus_2010, Graves_JLSGould_Serpens_2010}
which is equivalent to $\sim$0.4\,pc at a distance of 5.7\,kpc.

For the purpose of examining the column density and temperature of the region,
we retrieved the publicly available Vialactea \emph{Herschel}
column density and temperature
maps for the
same\footnote{\href{http://www.astro.cardiff.ac.uk/research/ViaLactea/}{http://www.astro.cardiff.ac.uk/research/ViaLactea/}}
\citep{Molinari_Vialactea_2010}. These maps have been generated by applying
the Bayesian point-process procedure \citep{Marsh_PPMAP_2015} to Hi-Gal survey
images \citep{Marsh_PPMAP_Higal_2017}.
The pixel scale of $\sim$6\arcsec\, and a resolution of $\sim$12\arcsec\, makes
them suitable to examine in conjunction with the JCMT data.
Finally, \emph{Herschel} PACS
\citep[Photodetector Array Camera and Spectrometer;][]{Poglitsch_PACS_2010AA}
70\,\micron\, image and SPIRE
\citep[Spectral and Photometric Imaging Receiver;][]{GriffinHerschelSPIRE_10}
250\, and 350\,\micron\, images
(Proposal : \textquoteleft OT2\_smolinar\_7\textquoteright)
were retrieved from the archives for morphological examination and
filament identification purposes.

\begin{table*}
\caption{Data Used in this work}
\label{table_dataused}
\begin{tabular}{lllll}
\hline
Data Source  &  Line/Wavelength   &  Spatial     & Channel  &  Reference   \\
             &                    &  Resolution  & Width    &              \\
\hline
\multicolumn{5}{c}{Spectral Data Products} \\
\hline
CGPS          &  CO(1--0)       &  $\sim$100.44\arcsec & $\sim$0.82\kms  & \citet{Taylor_2003_CGPS}  \\
\cmidrule(lr){2-5}
JCMT Archive  &  \co            &  $\sim$14\arcsec     & $\sim$0.42\kms  & \citet{Buckle_HARP_ACSIS_2009}  \\
\cmidrule(lr){2-5}
JCMT Archive  &  \tco, \cetno   &  $\sim$14\arcsec     & $\sim$0.05\kms  & \citet{Buckle_HARP_ACSIS_2009}  \\
              &                 &                      & (0.5\kms\, for &   \\
              &                 &                      & ~rebinned cubes)      &   \\
\hline
\multicolumn{5}{c}{Imaging Products} \\
\hline
Vialactea Maps  &     --        &  $\sim$12\arcsec     & --              & \citet{Molinari_Vialactea_2010}  \\
\emph{Herschel} Archive  &  70\,\micron &  $\sim$5\arcsec & -- & \citet{Poglitsch_PACS_2010AA} \\
                         &  250\,\micron, 350\,\micron    &  $\sim$18\arcsec,$\sim$25\arcsec & -- & \citet{GriffinHerschelSPIRE_10} \\
NVSS                     &  1.4\,GHz                    &  $\sim$45\arcsec\, & --  & \citet{Condon_NVSS_AJ_1998} \\
BGPS                     &  1.1\,mm                     &  $\sim$33\arcsec\, & --  & \citet{Aguirre_BGPS_ApJS_2011} \\
\hline
\end{tabular}
\end{table*}


\begin{figure*}
\centering
%
\renewcommand{\thesubfigure}{(left)}
\subfigure
{
\includegraphics[height=3.0in,width=3.0in]{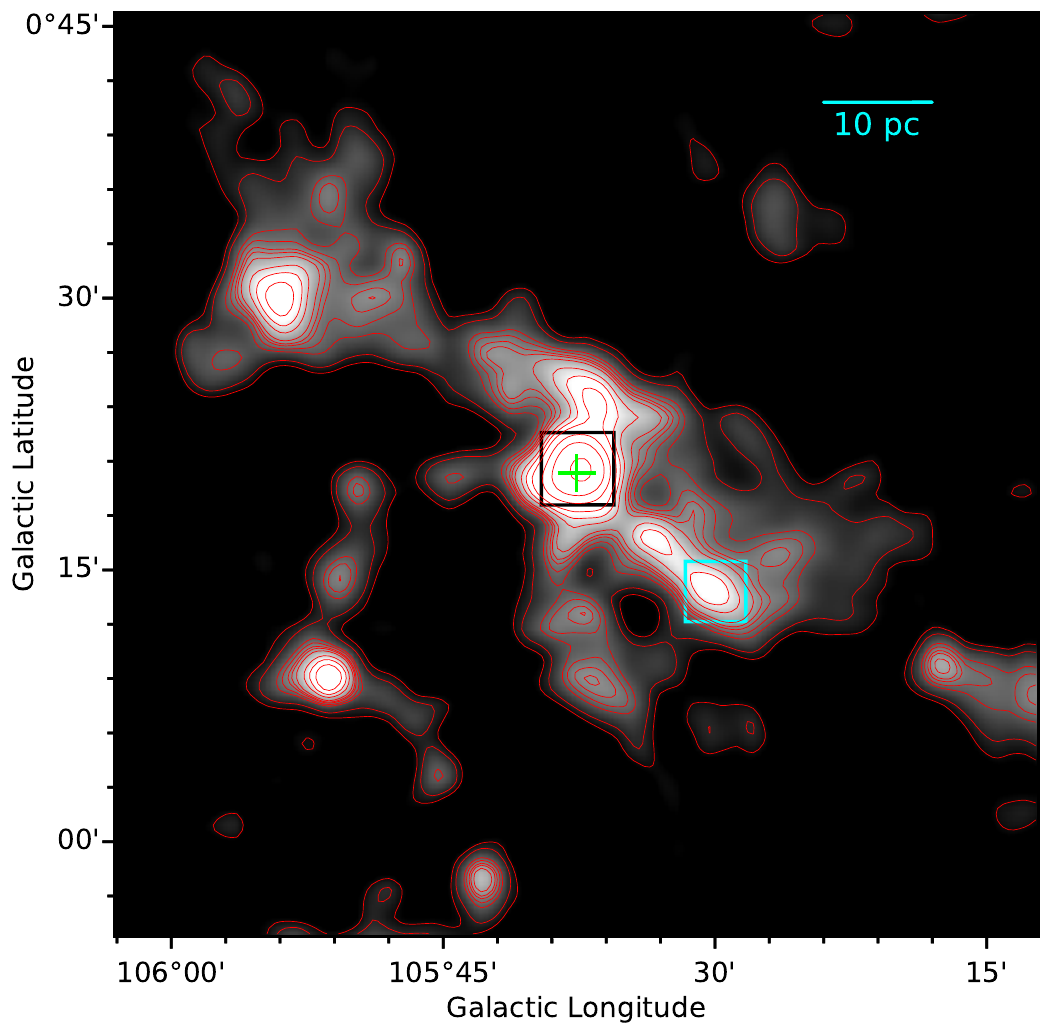}
\label{fig_cgps_m0_total}
}
\renewcommand{\thesubfigure}{(right)}
\subfigure
{
\includegraphics[height=3.0in,width=3.0in]{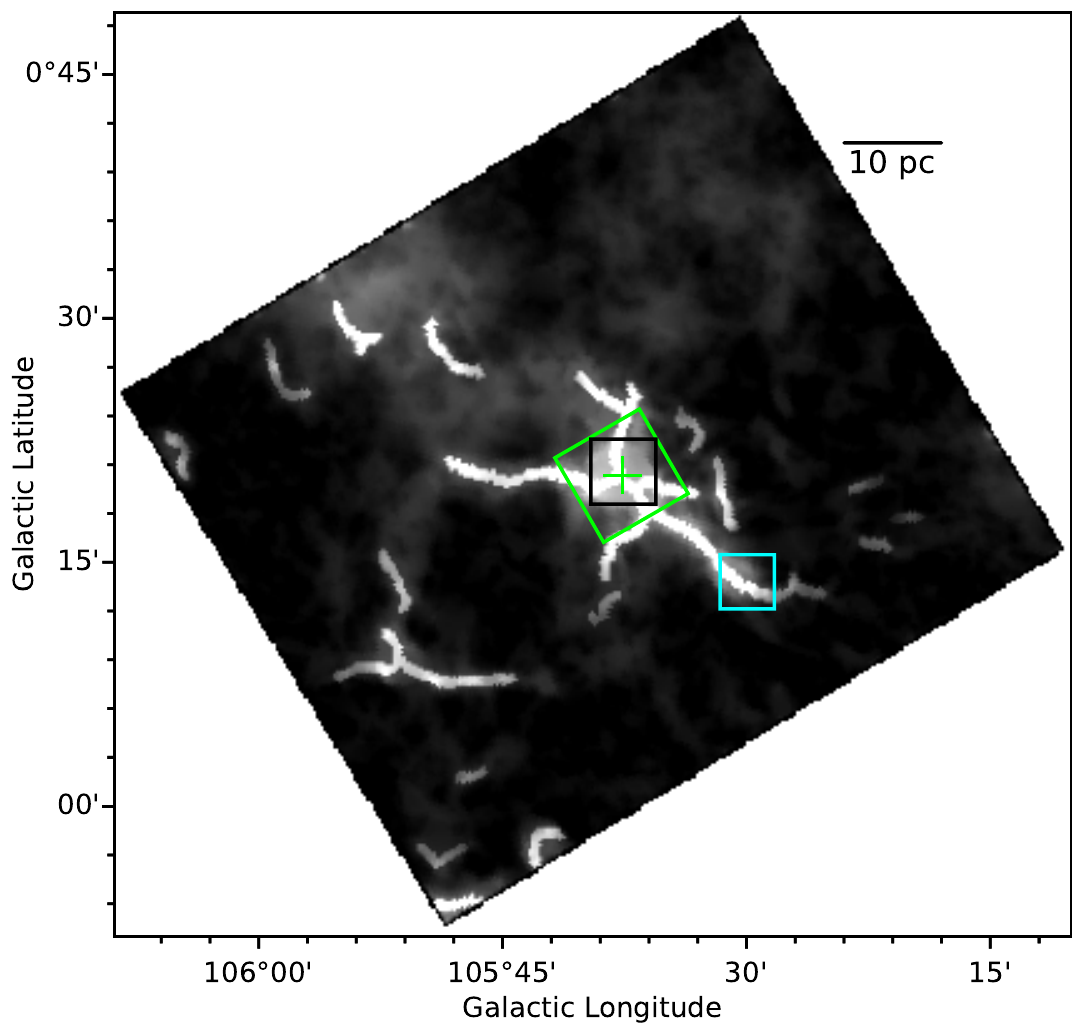}
\label{fig_350getsf}
}
\caption{
\emph{(left)}
CO(1-0) integrated intensity map in [-62.5,-41]\kms\, velocity range.
Contours have also been drawn (at 5, 7, 11, 12.5, 14, 15, 16.5, 20, 23, 27, 35, 45,
and 65 K\,km\,s$^{-1}$) for clarity of the features.
\emph{(right)}
\emph{Herschel} 350\,\micron\, image with overlaid skeletons of filaments
identified by \emph{getsf}.
Black Box denotes the central portion where CO(1--0) spectrum has been
extracted.
Cyan box marks the CO clump which also (i.e. apart from the central region)
corresponds to N(H$_2$) $\geq$ 10$^{22}$\,cm$^{-2}$
(see Figure \ref{fig_350getsf_central}).
Green plus symbol shows the location of IRAS\,22308+5812, and green box on the
350\,\micron\, image shows the field of view of JCMT analysis
(see Section \ref{section_kinematics_JCMTobs}).}
\label{fig_cgpsCO_350getsf}
\end{figure*}

\begin{figure*}
\centering
\includegraphics{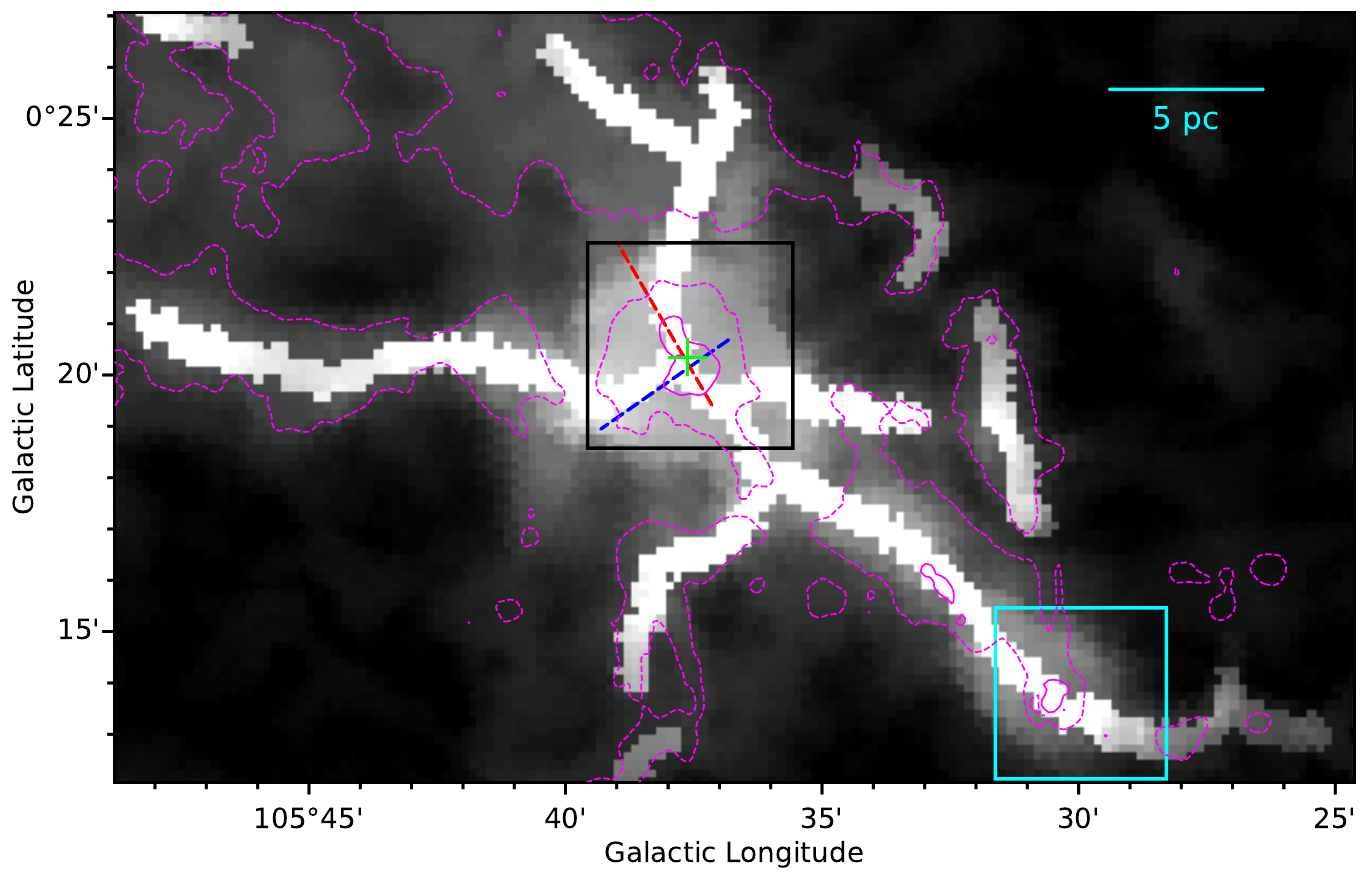}
\caption{A zoomed in view of Figure \ref{fig_350getsf}.
Magenta dashed and solid contours mark N(H$_2$) = 33$\times$10$^{20}$\,cm$^{-2}$
and 10$^{22}$\,cm$^{-2}$, respectively, from the Vialactea maps. Dashed blue and
red lines mark the major axes of blue and red outflow contours, respectively,
from \citet{Qin_wrtS138_2008}.
The rest of the symbols are same as Figure \ref{fig_cgpsCO_350getsf}.}
\label{fig_350getsf_central}
\end{figure*}

\begin{figure}
\includegraphics[width=\linewidth]{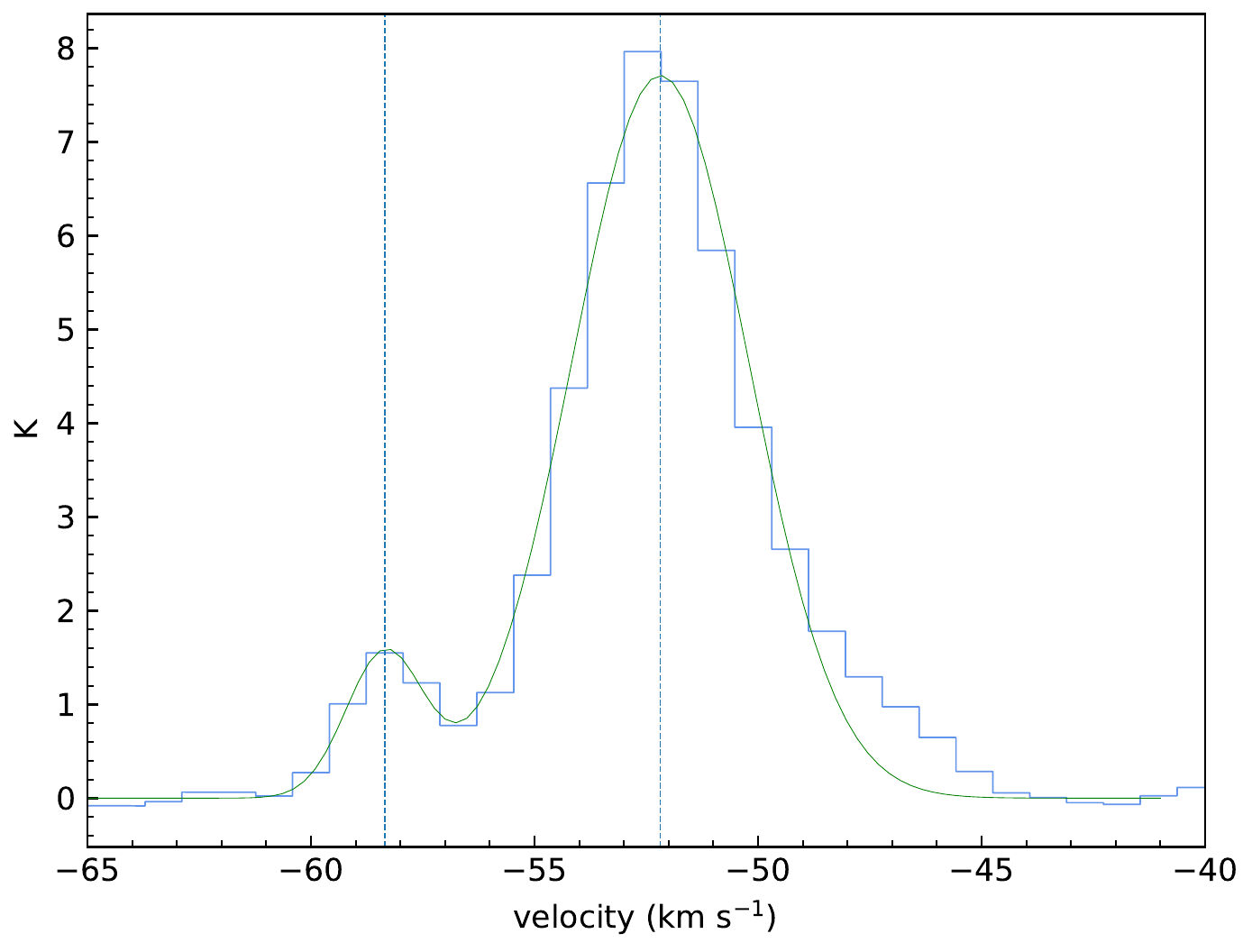}
\caption{
CO(1-0) spectrum at the central position marked in
Figure \ref{fig_cgps_m0_total}. Green curve shows the gaussian fit,
with dashed vertical lines marking the peak velocities of the two
velocity components.
}
\label{fig_cgps_spectra}
\end{figure}

\section{Analysis and Results}
\label{section_AnalysisResults}

\subsection{Large-scale view of the region}
\label{section_largescaleview}

In this section, we explore the large-scale morphology as well as
the hub-filament system in the region.
Figure \ref{fig_cgps_m0_total} shows the $^{12}$CO(1-0) moment-0
(integrated intensity) map of the larger region in the velocity range
[-62.5,-41]\kms.
The central core has been marked in a black box and the IRAS source
(IRAS\,22308+5812) with a green plus symbol. Multiple filamentary
structures can be seen emanating from this central core, with each of
them harbouring separate potential clumps of their own.
To better understand the filamentary structure of the region,
we used the \emph{getsf} tool (version 211109) of
\citet{Menshchikov_getsf_2021}.
The tool, especially developed for \emph{Herschel} images,
decomposes an image into its structural components (sources and filaments)
and separates them from their background.
The method is fully automated and takes as input only one parameter
from the user -- the maximum size of the structure to extract.
For our purpose, we used the \emph{Herschel} 350\,\micron\, image, and
set the maximum size of filamentary structure to 550\arcsec\,
based on visual inspection.
Figure \ref{fig_350getsf} shows the \emph{Herschel} 350\,\micron\,
image with the skeleton of filamentary structures identified by the
\emph{getsf} tool. It should be noted that the width associated with
the skeletons on the image is only for visualisation purpose.
It can be seen that the structures which appeared
nearly contiguous in emission
in the CO map (Figure \ref{fig_cgps_m0_total}), are detected as different
(filamentary) structures in Figure \ref{fig_350getsf}.
Figure \ref{fig_350getsf_central} shows a zoomed in view of Figure
\ref{fig_350getsf} with overlaid column density contours from Vialactea
map and the major axes of red and blue outflow lobes from \citet{Qin_wrtS138_2008}.
According to the criteria of \citet{MyersFilaments_09},
the hub has low aspect ratio, and
can be defined in terms of column density as the region
where N(H$_2$)\,$\gtrsim$\,10$^{22}$\,cm$^{-2}$.
As per this definition,
the hub region here (solid magenta contour in
Figure \ref{fig_350getsf_central}) occurs at the location where there
is a joining of multiple filaments. Some of the \emph{getsf} filaments
are also traced by lower column density contour
at 33$\times$10$^{20}$\,cm$^{-2}$.
However, the large scale view of the region shows that the filament sizes
have an order of magnitude of $\sim$\,10\,pc.
This is in contrast with nearby HFS, where filaments have been
found to have sizes of the order of magnitude of $\sim$ 1\,pc
\citep{Arzoumanian_2017_arXiv, Arzoumanian_nearbyfilaments_2019_AA}.
On the other hand, large filaments of sizes in tens of parsecs have also
been recorded in literature
\citep{Zucker_LargeScaleFil_2018_ApJ, Hacar_PP7_2022arXiv}.
Nevertheless, as also discussed in \citet{Nanda_2020},
we would like to add the caveat that higher resolution studies of the region
could resolve it into structures with similar hub/filament size scales
as for the nearby regions.

Figure \ref{fig_cgps_spectra} shows the $^{12}$CO(1-0) line spectrum
in the central region and the gaussian fit to its components.
The first velocity component is at ($v\pm\sigma$)
-58.4$\pm$0.9\,km\,s$^{-1}$ with an amplitude of 1.5\,K;
while the second component is at
-52.2$\pm$2.0\,km\,s$^{-1}$ with an amplitude of 7.7\,K.
These two different velocity components have been observed in earlier studies
as well \citep{Wouterloot_WB89_1989, Kerton_wrtS138_2003, Brunt_wrtS138_2003}.
It must be noted that CO(1-0) emission is regarded as optically thick and
often self-absorbed. Hence it is not used as a probe for the denser regions
of the molecular cloud.
For the central region, the stronger velocity component
has a prominent broad wing towards
the red side, which is probably an effect of the outflow.

\begin{figure*}
\centering
\includegraphics{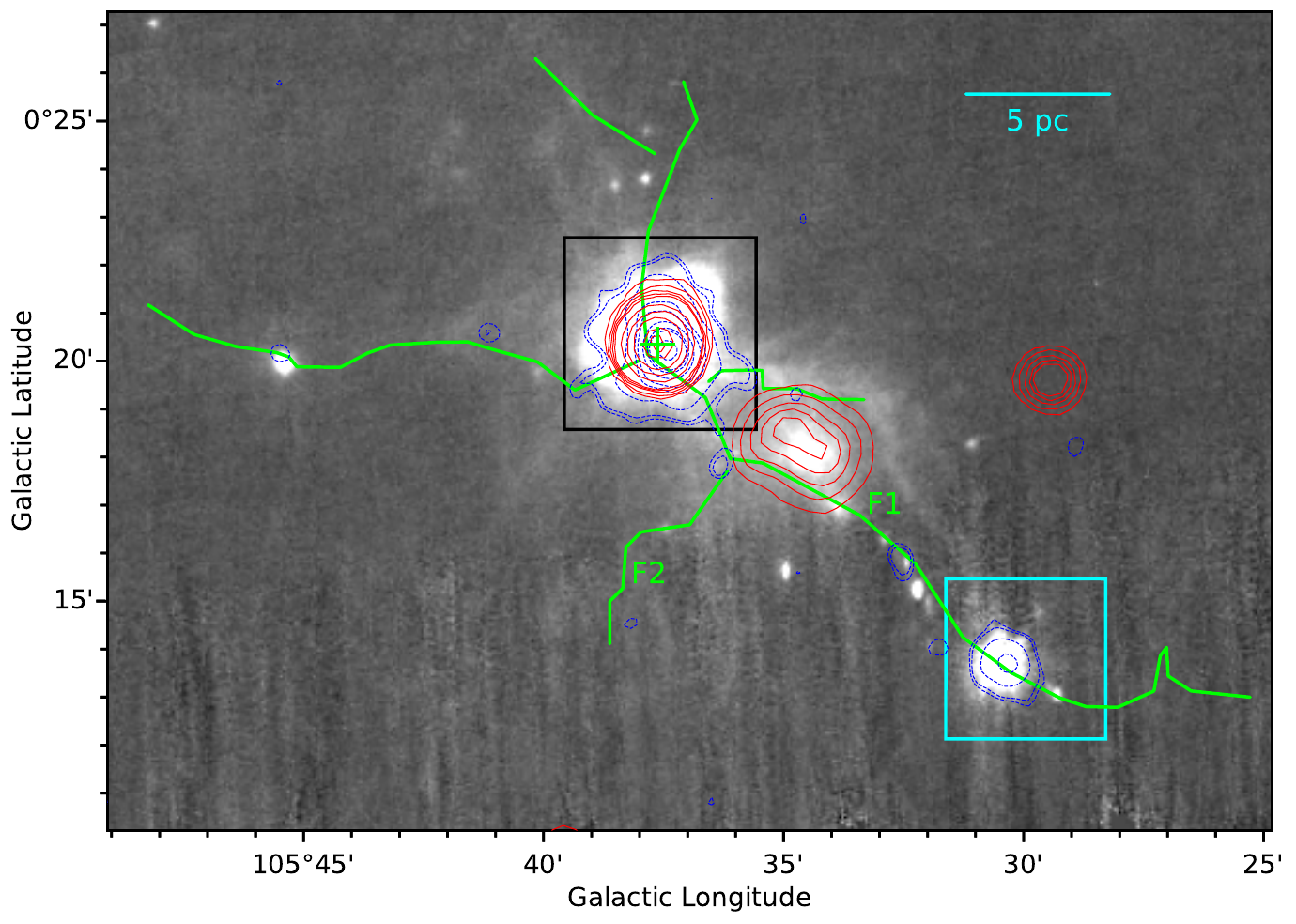}
\caption{
\emph{Herschel} 70\micron\, image of the hub-filament system.
Bolocam 1.1\,mm emission is shown in blue dashed contours at
0.08, 0.1, 0.2, 0.4, 0.5, 0.7, 0.9, 1.0, and 1.4 Jy\,beam$^{-1}$.
Red contours (at 0.002, 0.004, 0.007, 0.01, 0.013, 0.04, 0.08, 0.15, 0.3,
and 0.4 Jy\,beam$^{-1}$) are for the NVSS 1.4 GHz emission.
Green lines mark the \emph{getsf} filamentary skeletons
(see Figure \ref{fig_350getsf_central}).
Green cross shows the location of the IRAS source, and
``F1'' and ``F2'' are the labels for the two respective filaments.
The central black box and the cyan box are the same as
in Figure \ref{fig_350getsf_central}.}
\label{fig_ratio_map}
\end{figure*}

Figure \ref{fig_ratio_map} shows the \emph{Herschel} 70\,\micron\,
image of the region, overlaid with filament skeletons from the
\emph{getsf} tool (Figure \ref{fig_350getsf_central}), smoothed
Bolocam 1.1 mm contours (in blue), and NVSS 1.4 GHz contours (in red).
70\,\micron\, emission traces the warm dust emission due to stellar
sources and is indicative of star formation in a region
\citep{Li_70micron_2010ApJ,Calzetti_SFR_2013seg}.
It can be seen that along the filament marked ``F1'' on the image,
there appears to be star formation activity, as evidenced
by the presence of bright diffused emission along its length.
To the slight west of the point where the filament F1 joins filament F2,
extended emission can be traced on the image, also seen in NVSS (red
contours).
The integrated flux density from the NVSS image was obtained to be 0.06\,Jy,
yielding a Lyman continuum flux of $\sim$\,10$^{47.2}$\,photons\,s$^{-1}$
\citep{Moran_Radio_83},
which on comparison with the tabulated values from \citet{Panagia_73}
(assuming a Zero Age Main Sequence (ZAMS) single source)
would suggest that a source of at least spectral type B0.5--B0
is required for the ionisation of the region.
Though no massive stars have been found located in the immediate vicinity
of this diffused ionised region, it is possible that there could be embedded
massive star formation at the junction of filaments F1 and F2.
The presence of a 1.1 mm clump at this junction could be an affirmation of
this.

The filament F1 also hosts a few Bolocam 1.1 mm emission clumps along
its length, especially at its south-west corner, where there appears
to be a massive clump (within the cyan box).
We retrieved the integrated flux density from the
Bolocam Galactic Plane Survey Catalog v2.1 \citep{Ginsburg_Bolocam_ApJS_2013}
from the IRSA\footnote{https://irsa.ipac.caltech.edu/} archive.
Apart from the 1.1 mm emission in the central region (associated with
IRAS\,22308+5812), the catalog contains only the clump in the south-west
in our field of view.
Other contours depicting 1.1 mm emission might not have been included due
to not meeting the catalog criteria \citep{Ginsburg_Bolocam_ApJS_2013}.
Nevertheless, it is relevant to note their presence along the filaments.
The retrieved integrated flux density was used to calculate the core total mass
of gas and dust using the following formula
\citep{Hildebrand_MassFormula_QJRAS_1983,
Enoch_ApJ_2008, Bally_Bolocam_ApJ_2010} :

\begin{eqnarray}
M = \frac{D^2~S_{\nu}~R_t}{B_{\nu}(T_d)~\kappa_{\nu}},
\end{eqnarray}

where,
D is the distance,
S$_\nu$ is the integrated flux density,
R$_\textup{t}$ is the gas-to-dust mass ratio (taken as 100),
B$_{\nu}$(T$_\textup{d}$) is the Planck function at dust temperature
T$_\textup{d}$ (taken as 10\,K), and
$\kappa_{\nu}$ is the dust opacity (1.14\,cm$^2$\,g$^{-1}$)
\citep{Enoch_BolocamPerseus_ApJ_2006, Lokesh_S235_ApJ_2016}.
The above equation assumes that the emission (at 1.1\,mm here)
is optically thin, and both T$_\textup{d}$ and $\kappa_\textup{1.1\,mm}$ are
position-independent within a core.
Using the retrieved integrated flux densities of $\sim$\,8.316\,Jy
and 1.300\,Jy for the central and south-western clumps, respectively,
the resultant masses were calculated to be
$\sim$ 10274\,M$_\sun$ and 1606\,M$_\sun$, respectively.
The south-west Bolocam clump is also associated with a ``hub'', as is evident
in Figure \ref{fig_350getsf_central} from the column density contour
(solid magenta contour within the cyan box).
We find that while the south-west Bolocam clump has mass of the order of
magnitude similar to other hubs from literature
(i.e. $\sim$10$^3$\,M$_{\sun}$), such as
Mon\,R2 \citep{TrevinoMorales_MonR2_2019, Nanda_MonR2_2022AA},
SDC13 \citep{Williams_SDC13_2018AA}, and others \citep{Hacar_PP7_2022arXiv},
the mass for the central Bolocam clump is an order of magnitude larger.
This is probably due to the fact that the integrated flux density covers a
wider emission region \citep{Rosolowsky_BGPS_2010ApJS} than the mere hub.
If one were to use the flux density within 80\arcsec\, diameter aperture
for the central clump (area roughly coincident with
N(H$_2$)$\sim$10$^{22}$\,cm$^{-2}$ hub region)
-- also given in the catalog ($\sim$3.814\,Jy) --
a mass of $\sim$\,4712\,M$_{\sun}$ is obtained,
which is of the order of magnitude as for other studies in literature.

\begin{figure*}
\includegraphics[width=\textwidth]{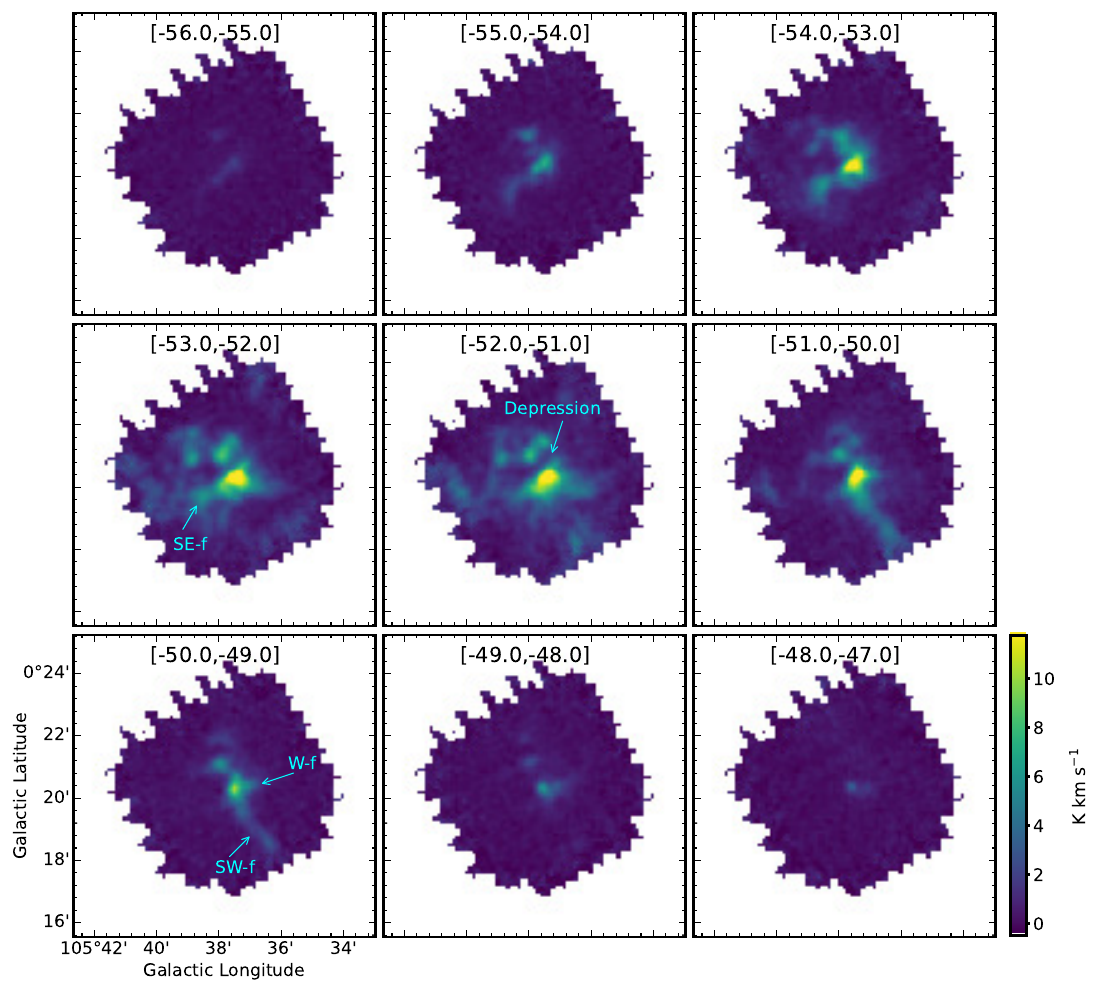}
\caption{Channel maps of \tco\, emission. Three filamentary structures --
Western filament (W-f), South-West filament (SW-f), and South-East filament (SE-f)
can be traced on the images. A region of minima in emission has been marked as
``Depression'' in the [-52,-51]\kms\, channel map.}
\label{fig_chmap_13CO32}
\end{figure*}

\subsection{Molecular gas kinematics in the central region}
\label{section_kinematics_JCMTobs}

We now examine the central part of the Sh2-138 region
(see the field of view marked on Figure \ref{fig_350getsf})
using (spatial and spectral) higher resolution \co, \tco, and
\cetno\, molecular line data retrieved from the JCMT archive.
The field of view of each cube is $\sim$\,5\arcmin$\times$5\arcmin\,
and roughly encloses the hub and central core of CGPS CO emission.
For $^{13}$CO and C$^{18}$O,
the rebinned cubes with 0.5\kms\, channel width have been used to
examine the spatial structures in the following sections as their
lower noise (see section \ref{section_dataused}) allows for better
examination of features.
However for the calculation of physical parameters (in section
\ref{section_jcmt_physicalparameters}), the native
channel width ($\sim$\,0.05\kms) $^{13}$CO and C$^{18}$O cubes have
been utilised as we need high velocity resolution here.


Figure \ref{fig_chmap_13CO32} shows the channel maps for the \tco\, emission.
The velocity range for each frame has been given. In the [-50,-49]\kms\,
velocity range,
two prominent filamentary structures can be traced, and these have been
indicated by arrows. The filament along the south-west direction (SW-f)
is longer and much more prominent.
It can only be seen in the -51 to -49\kms\, velocity
range, and has no counterparts at larger or smaller velocities.
The second filament, pointing west (W-f), is also seen prominently
in -51 to -49\kms\,,
but can be traced as diffused emission at larger and smaller velocities.
As we move towards higher (absolute) velocities, a third filament can
be traced in the -56 to -51\kms\, range in south-east direction (SE-f), and
is most prominently seen in the [-53,-52]\kms\, channel map.
Apart from the filaments, another prominent feature is the sudden depression
in emission - along the south-east to north-west axis - seen in the
[-53,-52] and [-52,-51]\kms\, channel maps.
At larger and smaller velocity channel maps,
this depression shows the presence of diffuse emission.
It should be noted that the systemic velocity of the cloud complex is
$\sim$\,-51.7$\pm$1.9\,\kms\,(see Section \ref{section_jcmt_physicalparameters}).
We note that the above-mentioned features (filaments and depression) could
also be traced in the \cetno\, channels, albeit not at the requisite
signal-to-noise threshold (see section \ref{section_jcmt_momentmaps}).

\begin{figure*}
\includegraphics[width=\textwidth]{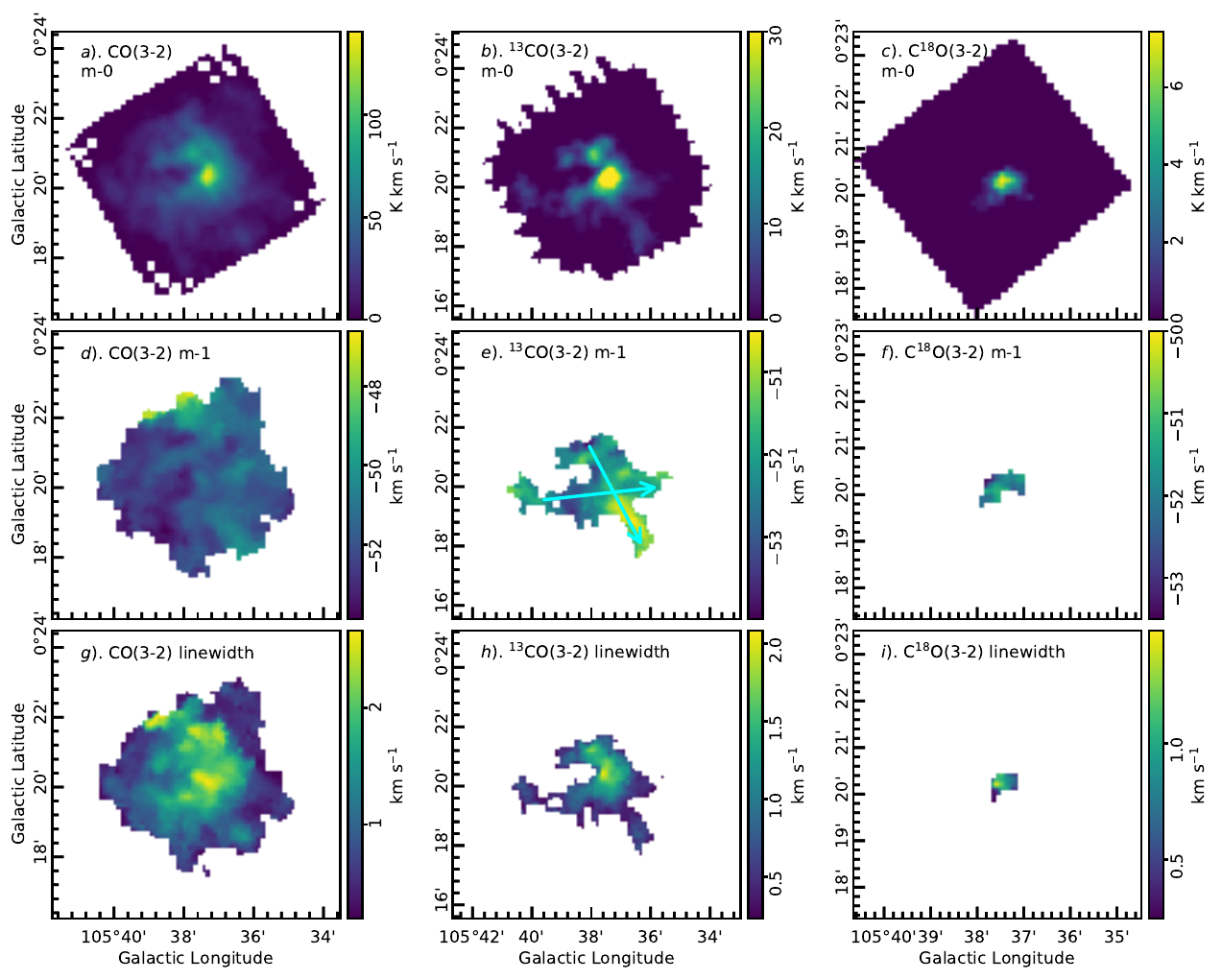}
\caption{Row-wise :
moment-0 (Integrated intensity),
moment-1 (Intensity-weighted velocity), and
linewidth (Intensity-weighted dispersion)
collapsed images for three cubes -- \co, \tco, \cetno\, in first, second, and
third columns, respectively.
Cyan vectors in \tco\, moment-1 map mark the directions along which p-v slices
have been extracted.}
\label{fig_m012_combined}
\end{figure*}

~\\
\subsubsection{Moment maps}
\label{section_jcmt_momentmaps}

To account for the noise in the cubes, we confine our further analysis
to only those regions with detections above 5$\sigma$ (where $\sigma$ is
the data's rms noise level).
Using the clumpfind algorithm \citep{Williams_clumpfind_94} as
implemented in the \textsc{cupid} package \citep{Berry_StarlinkCupid_2007}
of \textsc{starlink} software suite \citep{Currie_starlink_2014}, clumps
were identified in the position-position-velocity spectral cube.
The threshold for detection was kept conservatively at 5$\sigma$ to avoid
the possibility of false detections, and the gap between contour levels
at 2$\sigma$ as is recommended by \citet{Williams_clumpfind_94}.
The clumps detected in all three
spectral cubes (\co, \tco, and \cetno) lay in the range (-60,-45)\kms.
Thereafter, the regions with non-detection were masked to construct
masked cubes for all three molecular lines.

Figure \ref{fig_m012_combined} shows the moment-0 (Integrated emission),
moment-1 (Intensity-weighted velocity), and linewidth (Intensity-weighted
dispersion) collapsed images from all three (masked) cubes.
In the intensity-weighted velocity map for \co,
high-velocity gas is interspersed
with low-velocity gas throughout and no particular pattern is discernible.
For \tco, as in the case of integrated intensity emission, filamentary structure
is noticeable here as well. The filament along the south-west
(SW-f, see Figure \ref{fig_chmap_13CO32}) shows a lower
(absolute) velocity than other parts of the region.
Such a velocity profile would suggest a gradient (at the joint between the SW-f
filament and the central part) resulting in
gas flow. To better understand the same, position-velocity
slices were taken along the two vector directions (east-west and north-south)
marked on this image,
which we elaborate upon in section \ref{section_jcmt_pv}.

The intensity-weighted dispersion maps are shown in the last row of this
image grid.
The central part of the cloud shows a large velocity dispersion, of the order
of about 2\kms\, in all three molecular lines, yet another indication of
mixing of flows.
This large dispersion is seen to extend along the north-east part of the
image in both \co\, and \tco\, molecular lines.
In contrast, the south-west filament (traced prominently in \tco\,) shows a
relatively smaller dispersion of $\sim$1\kms.
\co\, is the most
ubiquitous tracer of molecular hydrogen and shows diffused emission in the
entire region in the integrated intensity map.
\tco\, has higher critical density and thus shows the filamentary
nature prominently, while \cetno\, which has the highest critical density
primarily traces the central clump.

\begin{figure*}
\includegraphics[width=\textwidth]{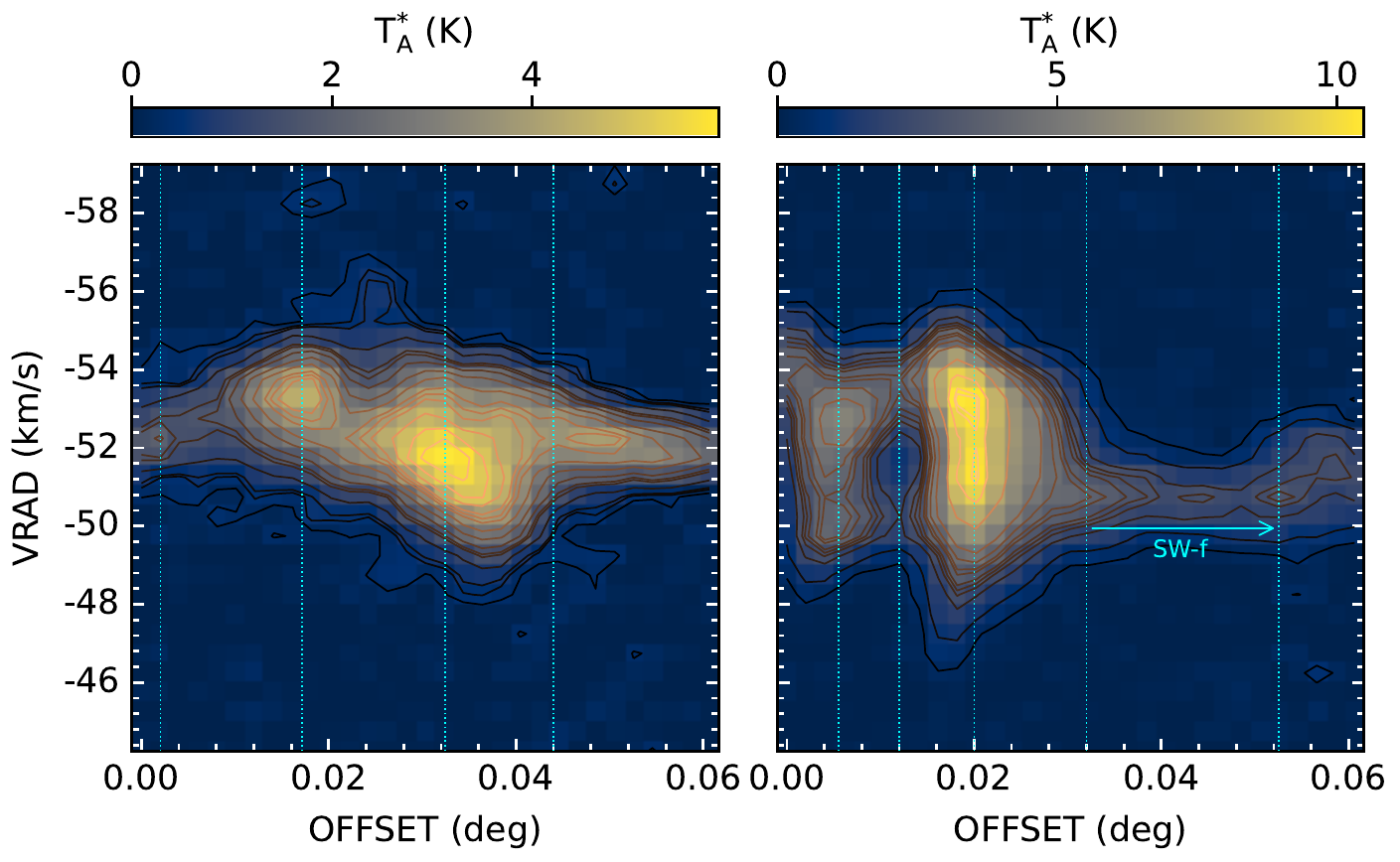}
\caption{
Position-velocity maps along the two vector
directions marked in the \tco\, intensity-weighted velocity map in
Figure \ref{fig_m012_combined}(e).
Left : along the east-west vector
(contours at 0.3, 0.5, 0.6, 1, 1.5, 1.65, 2, 2.5, 3, 3.5, 3.75, 4, 4.5, 5, 5.5, and 6 K).
Right : along the north-south vector
(contours at 0.45, 1, 2, 2.5, 2.75, 3, 3.5, 3.75, 4.25, 5, 7, 9, 10, and 10.25 K).
Vertical dotted lines mark the intervals between which appreciable gradient
can be seen. SW-f is the filament from Figure \ref{fig_chmap_13CO32}.}
\label{fig_13co_pv}
\end{figure*}

\subsubsection{Position-velocity slices}
\label{section_jcmt_pv}

Figure \ref{fig_13co_pv} shows the position-velocity (p-v) maps along the two
directions marked in the \tco\, intensity-weighted velocity map in
Figure \ref{fig_m012_combined}(e).
The first figure shows the slice for the east-west vector and second for the
north-south vector.
Along both the slices, there appears to be a significant velocity gradient
as one moves along the vector.
Vertical dotted lines mark the intervals between which appreciable gradient
can be seen.
The east-west slice shows the absolute velocity increasing from
$\sim$ 52\kms\, to 54\kms, then decreasing to below 51\kms\, and then
again increasing to above 52\kms.
A similar pattern is observed in the north-south slice, where the velocity
first increases slightly and then decreases as one moves along the vector,
i.e. from offset 0 degrees to 0.06 degrees ($\sim$\,6\,pc).
In Figure \ref{fig_13co_pv}(left),
the offset from $\sim$\,0.044 degree onwards is roughly coincident with
the W-f filament, and in Figure \ref{fig_13co_pv}(right),
the offset from $\sim$\,0.032--0.052 degrees represents the
SW-f filament (see \tco\, moment-1 map in Figure \ref{fig_m012_combined}(e)).
As was also seen in the moment-1 map, the velocity is nearly constant
at $\sim$\,-52\kms\, and -51\kms\, along W-f and SW-f filaments, respectively.
A small gradient is seen only towards the end for SW-f.
This gradient coincides with the position of peak(s)/clump(s)
(discussed below in section \ref{section_jcmt_spectra})
and could thus denote gas being channeled into the clump.
Overall, this gradient pattern in both the slices is indicative of gas
spiralling into the central region from both the ends of the molecular cloud.
There also appears to be clumping of gas, with
the brightest clump (roughly at an offset of 0.032 degrees and 0.02 degrees in
figures \ref{fig_13co_pv}(left) and \ref{fig_13co_pv}(right) respectively)
being that of the central region.
Another noticeable feature in the north-south slice (Figure \ref{fig_13co_pv}(right))
is the presence of a dark region
at an offset of $\sim$ 0.012 degrees in the velocity range -52 to -51\kms.
There is only a faint connecting feature between the immediate northern and
southern clumps at $\sim$ -54 to -53\kms.
This dark region corresponds to the prominent gap seen in the channel map
(marked ``Depression'' in the [-52, -51]\kms\, channel map of Figure \ref{fig_chmap_13CO32}).
In the central region, we find material clumping at two distinct
velocities -- -51 and -53\kms.
Overall, the maps show considerable dispersion in the velocity space around
the systemic velocity of the cloud complex ($\sim$\,-51.7\kms\,), possibly
due to the two directions being along the two outflow axes in the
region (see section \ref{section_discussion} for further discussion).

\subsubsection{Analysis of \tco\, and Herschel clumps}
\label{section_jcmt_spectra}

\begin{figure*}
\includegraphics[width=\textwidth]{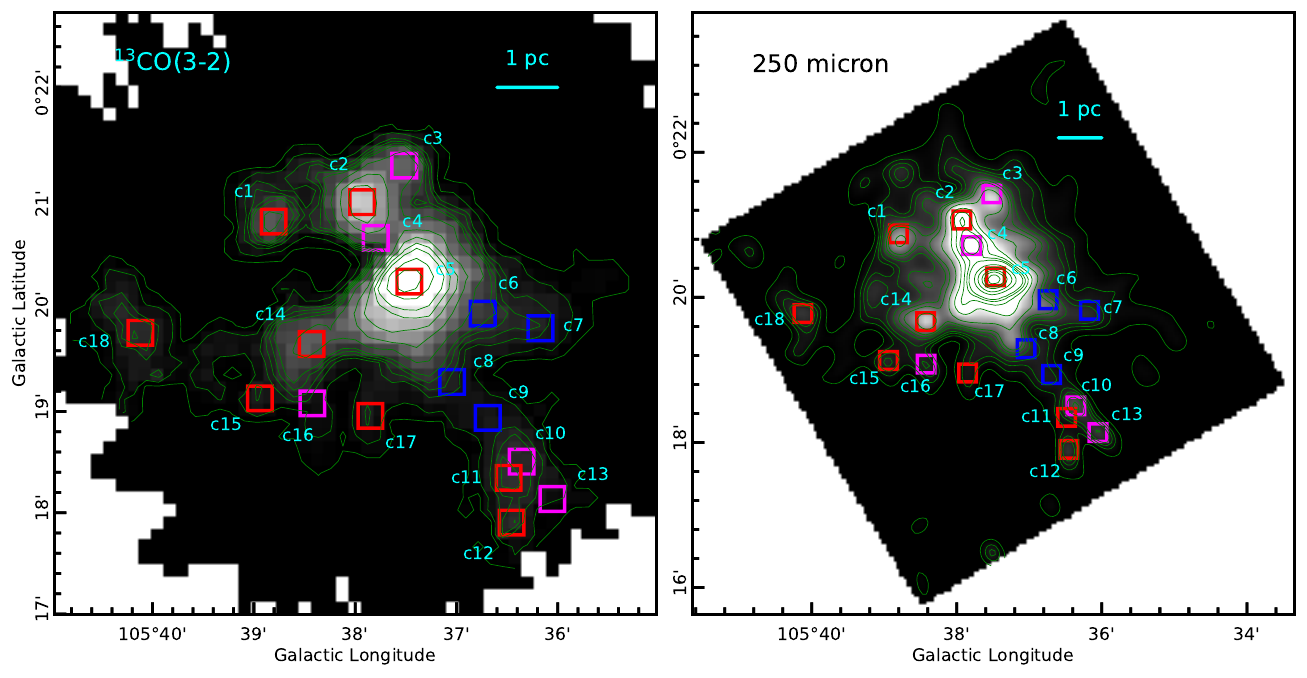}
\caption{
Left : \tco\, integrated intensity map
(contours at 0.45, 2, 4, 5, 6, 7, 10, 15, 20, 25, 30, 40, and 50 K\,km\,s$^{-1}$).
Right : \emph{Herschel} 250\,\micron\, image
(contours at 0.23, 0.3, 0.35, 0.4, 0.5, 0.6, 0.7, 1, 1.5, 2, 2.5, 3, 5, 7, 10,
12, and 15 $\times$10$^3$ MJy\,sr$^{-1}$).
Red boxes mark the peaks seen in the integrated intensity map while
magenta boxes mark those which are primarily seen in the 250\,\micron\, image (and are not
noticeable in the first image). Blue boxes mark the locations which do not display any peak,
but where spectra for
W-f and SW-f (i.e. western and south-west filaments, Figure \ref{fig_chmap_13CO32})
were extracted.}
\label{fig_13co_spire250}
\end{figure*}

\begin{figure*}
\includegraphics[width=\textwidth]{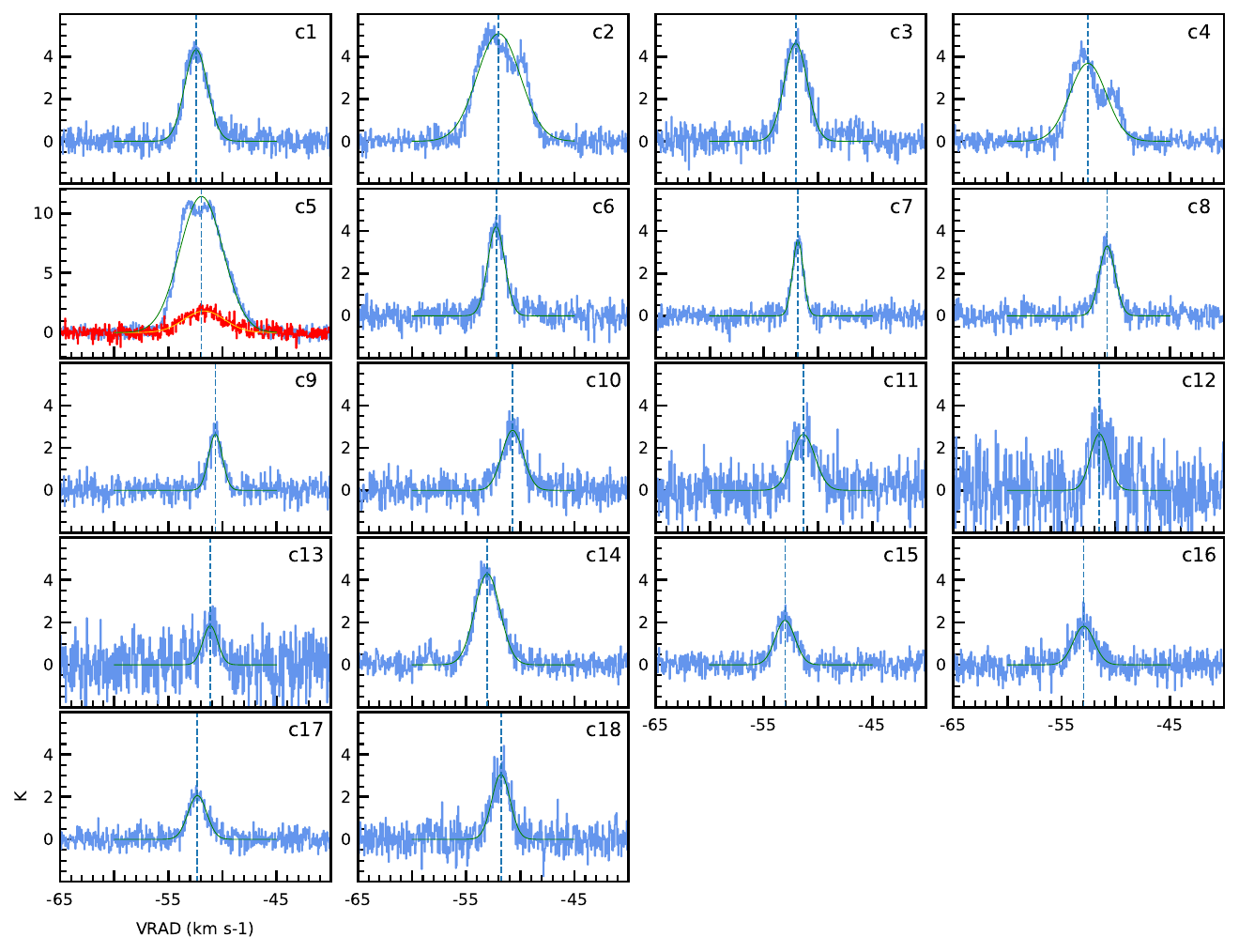}
\caption{\tco\, spectra at positions marked in Figure \ref{fig_13co_spire250}.
Green curve depicts the gaussian fit to the spectra, with blue dashed line
marking the velocity for the peak of the gaussian fit.
For c5, \cetno\, spectrum has been plotted in red and gaussian model fit shown
with yellow curve.}
\label{fig_13co_spectra}
\end{figure*}

Figure \ref{fig_13co_spire250} shows the \tco\, integrated emission
along with \emph{Herschel} 250\,\micron\, map of the region. Contours
have been drawn on both the images for a better clarity of the features.
Boxes on both the images mark the locations where spectra have been extracted.
Since the beamsize is of the order of $\sim$2\,pixels, spectra were extracted
for a 2x2 box encompassing the pixel with the local maxima.
Red boxes mark the regions which show a (local) peak in the \tco\,
integrated intensity map, while magenta boxes (c3, c4, c10, c13, c16) mark
the locations of those peaks which are seen in 250 \micron\, but not traced
clearly in the integrated intensity image.
Blue boxes mark the locations which do not display any peak, but are along
the filaments which can be delineated --
c6 and c7 for the west-pointing filament (W-f in Figure \ref{fig_chmap_13CO32}),
c8 and c9 for the south-west pointing filament (SW-f in Figure \ref{fig_chmap_13CO32}).
The \tco\, molecular line spectra extracted at these locations have been
shown in Figure \ref{fig_13co_spectra}. For the central clump -- marked c5 --
\cetno\, emission was also detected at above 5$\sigma$ threshold, and hence
\cetno\, molecular line spectrum has also been plotted for this location
in Figure \ref{fig_13co_spectra} (weaker step-line plot in the frame for c5).
It should be noted that the massive stars studied by
\citet{Deharveng_S138_1999, Tapas_S138_2015} are associated with the location
of c5.

An obvious feature to note here is that the locations c2, c4, and c5 show
self-absorption near the peaks. While the spectrum for c5 (the central clump)
is nearly symmetrical around the self-absorption ``dip'' in the spectrum,
those for c2 and c4 are asymmetrical. The location c4 particularly seems to be
having high self-absorption, indicative of dense material present in this
part. It is worth noting that while the c4 clump is prominently observed
in the thermal dust emission wavelength at 250 \micron, the location shows a
depression in \tco\, molecular line emission. This can be seen in the integrated
emission in Figure \ref{fig_13co_spire250}(left) as well as in the channel maps
in Figure \ref{fig_chmap_13CO32}. c4 lies in the prominent gap marked in
the [-52, -51]\kms\, channel map of Figure \ref{fig_chmap_13CO32}.
It is also almost coincident with the gap seen in the position-velocity slice
(see Figure \ref{fig_13co_pv}(right) at around an offset of 0.012 degrees) along
the north-south vector (see intensity-weighted velocity map of \tco\, in
Figure \ref{fig_m012_combined}(e)).
We note that the weaker velocity component from the CGPS CO(1-0) spectra
(see Figure \ref{fig_cgps_spectra}) is barely seen and only for a few peaks
(such as c8, c13, c14, c17, and c18) due
to the weak nature of this emission. The low number of pixels used to
extract spectra at these locations make this component virtually
indistinguishable from the noise. An averaged spectrum over the entire JCMT
field of view does indeed reveal this component (not shown here).

\subsubsection{Physical parameters}
\label{section_jcmt_physicalparameters}

To extract the parameters for further calculations, we carried out gaussian
model fitting for each region in the velocity range [-60,-45]\kms\,
on the (non-averaged) $^{13}$CO and C$^{18}$O spectral cubes with
native channel width of $\sim$\,0.05\kms.
The gaussian fits to the \tco\, molecular line emission have been shown with
a green line in all the spectra (Figure \ref{fig_13co_spectra}).
The results of the model fitting of the clumps, along with other parameters,
are listed in Table \ref{table_fitting}.
As mentioned earlier, for the c5 central clump, \cetno\, spectrum was also
available, and has been shown as well along with its gaussian fit.
Both the isotopologues were found to have similar peak velocities,
$\sim$\,-51.7$\pm$1.9\kms\, and -52$\pm$2\kms\, for C$^{18}$O and $^{13}$CO,
respectively.
Since C$^{18}$O traces the densest part of the cloud, its velocity can
be taken as the systemic velocity of the cloud complex, and is in agreement
with literature
\citep{Blitz_wrtS138_1982, Burov_wrtS138_1988, Harju_wrtS138_1993, Bronfman_wrtS138_1996,  Johansson_wrtS138_1994}.

The gaussian model fit was used to obtain the FWHM (full width at half maxima)
and standard deviation (or velocity dispersion) for each region.
Thereafter, we calculate the non-thermal
velocity dispersion and total velocity dispersion using the following equations
\citep{FullerAndMyers_1992, FiegeAndPudritz_2000} :

\begin{eqnarray}
\Delta V_{tot}^2 &=& \Delta V_{obs}^2 + 8~~ln\,2~~kT \left( \frac{1}{\bar{m}} - \frac{1}{m_{obs}} \right) \\
\Rightarrow \frac{\Delta V_{tot}^2}{8~~ln\,2} &=& \frac{kT}{\bar{m}} + \left( \frac{\Delta V_{obs}^2}{8~~ln\,2} - \frac{kT}{m_{obs}} \right) \nonumber \\
\Rightarrow \sigma_{tot}^2 &=& c_s^2 + \left( \sigma_{obs}^2 - \sigma_\textsc{t}^2 \right) \\
&=& c_s^2 + \sigma_\textsc{nt}^2
\end{eqnarray}

where, $\Delta V_{obs}$ and $\sigma_{obs}$ ($=\Delta V_{obs}/\sqrt{8ln\,2}~$ for
a gaussian) are
the FWHM and standard deviation (or dispersion), respectively,
from the observed spectrum of the molecular species;
$\sigma_\textsc{t} (=\sqrt{kT/m_{obs}})$ is the thermal velocity dispersion for the
molecular species;
$m_{obs}$ is the mass of the molecule (29 amu and 30 amu for $^{13}$CO and
C$^{18}$O respectively);
$\sigma_\textsc{nt}$ is the non-thermal velocity dispersion;
$c_s$($=\sqrt{kT/\bar{m}}$) is the speed of sound;
$\bar{m}$ is the average molecular weight of the medium (2.37 amu);
and T is the excitation or gas kinetic temperature.

The critical density for \tco\, is in the range $\sim$\,10$^{4-5}$\,cm$^{-3}$,
and if we assume the gas and dust temperatures to be coupled
via collisions at this density \citep{Goldsmith_DustGasTemp_ApJ_2001},
then the dust temperature from \emph{Herschel} Vialactea temperature
map can be used for gas kinetic temperature T in the above equation.
We add a caveat though that, depending on the physical conditions of the region,
a significant difference between the gas and dust temperatures can
still exist \citep{Banerjee_MNRAS_2006, Koumpia_DustGasTemp_S140_AA_2015}.
The median temperature at the locations of the peaks ranged from 18-20\,K
(c$_s$ $\sim$0.25--0.27\kms), with the c4 and c5 clumps at 28\,K
(c$_s$ $\sim$0.31\kms).
At these temperatures, the thermal velocity dispersion $\sigma_\textsc{t}$
(for both $^{13}$CO and C$^{18}$O) comes out to be $\sim$0.07--0.09\kms.
Table \ref{table_fitting} lists these calculated values for the clumps
associated with the peaks in Figure \ref{fig_13co_spire250}.

The FWHM values for almost all the peaks lie in the
$\sim$ 2-2.75\kms\, range (and thus $\sigma_{obs}\sim$0.85-1.2\kms),
except for c2, c4, and c5 which lie in $\sim$4-4.7\kms
($\sigma_{obs}\sim$1.7-2\kms).
This was also observed in the linewidth map in Figure \ref{fig_m012_combined}(h).
For c13, the lower value might be due to the noisy spectrum.
As $\sigma_{obs}$ is much larger than $\sigma_\textsc{t}$ (0.07--0.09\kms) for
all the cases, we find it to be almost same as $\sigma_\textsc{nt}$.
c2, c4, and c5 locations have the highest non-thermal dispersions.
Using the above variables, we subsequently calculated the Mach number
($= \sigma_\textsc{nt}/c_s$) and the ratio of thermal to non-thermal pressure
($P_\textsc{tnt} = c_s^2/\sigma_\textsc{nt}^2$) as well \citep{Lada_ApJ_2003}.
All the locations were found to have Mach numbers which suggest
supersonic motion, with c2 and c5 displaying the largest values
($\gtrsim$6).
The order of magnitude of P$_\textsc{tnt}$ ($\sim$0.01--0.1) indicates that
non-thermal pressure dominates in the cloud, and the locations of highest
non-thermal dispersion and Mach numbers (i.e. c2, c4, and c5) were found to be
correlated to the least ratio values.

\citet{MyersFilaments_09} posit a hub-filament model of star-forming complexes
where the hub \citep{Nanda_2020}
is traced at a high column density of $\sim$\,10$^{22}$\,cm$^{-2}$ as
opposed to the filaments \citep{Andre_HerschelViewSF_13, Arzoumanian_2013}
which have a column density of $\sim$\,10$^{21}$\,cm$^{-2}$.
For the three filaments -- SW-f, W-f, and SE-f (see Figure \ref{fig_chmap_13CO32})
-- we calculated the line mass using the \emph{Herschel} column density map
constructed from thermal dust emission (see Section \ref{section_dataused}).
The length of the three filaments was taken from the 10$^{22}$\,cm$^{-2}$ contour
(i.e. the inner extent) along the line joining c6 and c7 for W-f; c8, c9, and c10
for SW-f; and upto c14 for SE-f.
Total column density was summed up in a width of 12\arcsec\,
(corresponding to $\sim$\,0.33\,pc at a distance of 5.7\,kpc) as it is the resolution
of the column density map. Thereafter, using the following formula
\citep{Mallick16148_15, Lokesh_S237_2017} :

\begin{eqnarray}
M_{line,obs} = \frac{\mu_\textsc{h$_2$}~m_\textsc{h}~Area_{pixel}~\Sigma N(H_2)}{\textup{(length of filament)}}
\end{eqnarray}

-- where $\mu_\textsc{h$_2$}$ is the mean molecular weight (2.8),
m$_\textsc{h}$ is the mass of Hydrogen,
Area$_{pixel}$ is the area subtended by one pixel,
and $\Sigma N(H_2)$ is the total column density --
we calculated the (observed) line masses to be
$\sim$\,32, 33.5, and 50\,M$_{\sun}$\,pc$^{-1}$
for the filaments W-f, SW-f, and SE-f, respectively.

\begin{table*}
\caption{Parameters derived from \tco\, spectra at the locations of peaks in
         Figure \ref{fig_13co_spire250}.
         For c5, \cetno\, spectrum
         was also available and thus used for calculation as well.}
\label{table_fitting}
\begin{tabular}{llllllll}
\hline
clump     &       FWHM             &       T            &       $\sigma_\textsc{nt}$    &       Mach        &       $P_\textsc{tnt}$      \\
          &       ($km~s^{-1}$)    &       (K)          &       ($km~s^{-1}$)    &       Number      &                      \\
\hline
c1        &       2.48             &       20           &       1.05             &       3.97         &       0.06           \\
c2        &       4.71             &       20           &       2.00             &       7.55         &       0.02           \\
c3        &       2.50             &       20           &       1.06             &       3.99         &       0.06           \\
c4        &       3.98             &       28           &       1.69             &       5.39         &       0.03           \\
\hline
c5        &  &  &  &  &  \\
~~~C$^{18}$O   &       4.39             &       28           &       1.86             &       5.95         &       0.03           \\
~~~$^{13}$CO   &       4.69             &       28           &       1.99             &       6.35         &       0.03           \\
\hline
c10       &       2.30             &       18           &       0.98             &       3.88         &       0.07           \\
c11       &       2.53             &       18           &       1.07             &       4.26         &       0.05           \\
c12       &       1.97             &       18           &       0.83             &       3.31         &       0.09           \\
c13       &       1.60             &       18           &       0.68             &       2.69         &       0.14           \\
c14       &       2.75             &       20           &       1.17             &       4.40         &       0.05           \\
c15       &       2.14             &       18           &       0.91             &       3.61         &       0.08           \\
c16       &       2.27             &       18           &       0.96             &       3.83         &       0.07           \\
c17       &       2.03             &       18           &       0.86             &       3.42         &       0.09           \\
c18       &       1.99             &       18           &       0.84             &       3.35         &       0.09           \\
\hline
\end{tabular}
\end{table*}


\section{Discussion}
\label{section_discussion}

According to the multiwavelength study of \citet{Tapas_S138_2015},
Sh2-138 represents the archetype ``hub-filament'' structure of
\citet{MyersFilaments_09} which
-- after the advent of \emph{Herschel} far-infrared data -- has been
found to be a ubiquitous feature of young stellar clusters hosting
low-mass and high-mass stars \citep{Nanda_2020}.
However, this interpretation needs to be tempered by the fact that
most of the filamentary clouds whose intricate structure has been
studied in literature have been nearby regions such as
Monoceros R2 \citep[830\,pc]{TrevinoMorales_MonR2_2019},
W40 \citep[500\,pc]{MallickW40_13},
IC\,5146 \citep[460\,pc]{Arzoumanian_IC5146_2011AA},
and others
\citep{Arzoumanian_nearbyfilaments_2019_AA, Hacar_PP7_2022arXiv}.
Though regions at further distances have been explored in literature,
such as
IRAS 05480+2545 \citep[2.1\,kpc]{Lokesh_IRAS05480_2017},
Sh2-53 \citep[4\,kpc]{Tapas_S53_2018},
G18.88–0.49 \citep[5\,kpc]{Lokesh_G1888_2020} and so on, the larger distance often
makes it difficult to resolve finer structures.
The large scale view of the Sh2-138 region shows filaments whose
sizes are of the order of $\sim$\,10\,pc.
Such large filaments (at distances of a few kpc) have also been
discussed in various studies
-- with Nessie cloud often held up as an archetype;
and terminologies such as ``Giant Molecular Filaments'' and ``Milky Way Bone''
have been employed for such large filaments
\citep{Ragan_GMF_2014AA, Zucker_LargeScaleFil_2018_ApJ}.
Nevertheless, as also discussed in \citet{Nanda_2020}, it is
possible that higher resolution studies of distant regions could
resolve a large filament into structures with size scales as
for the nearby regions.
The higher resolution JCMT and \emph{Herschel} maps for the central
region seem to suggest this where we can see more detailed structures.
The filamentary structures in the central region have sizes of the
order of a few parsecs, which is not so uncommon.
Examples of some of the massive star forming regions from literature
where such length scales have been observed are
G22 \citep[3.51\,kpc]{Yuan_G22_2018ApJ},
DR21 \citep[1.4\,kpc]{Hennemann_DR21_2012AA},
SDC13 \citep[3.6\,kpc]{Peretto_SDC13_2014},
though the line masses for these regions are larger with a wide range,
varying from $\sim$\,2 upto 20 times our line masses.
On the other hand, nearby regions such as
IC\,5146 \citep[460\,pc]{Arzoumanian_IC5146_2011AA} show a wide span
of line masses for filaments of lengths of a few parsecs, from
$\sim$\,0.5 to 3 times our line masses.

\citet{Hacar_PP7_2022arXiv} have used a census of more than 22000
filaments from literature to categorise them into different
(non-mutually exclusive) families.
A comparison of the structure and physical parameters of the filamentary
structures in Sh2-138 with their compilation shows that while our
filaments are similar to ``Dense Fiber'' filament family in their
categorisation in terms of being
structures in position-position-velocity space,
our filaments display a larger non-thermal dispersion
($\sigma_\textsc{nt}$). Higher $\sigma_\textsc{nt}$ values are seen for
``Galactic plane survey filaments'' and ``Giant filaments''
\citep[see Table 2 of][]{Hacar_PP7_2022arXiv}. As discussed for these two
filament families in \citet{Hacar_PP7_2022arXiv}, the calculation of
physical parameters could suffer from sensitivity and resolution biases,
given the 5.7\,kpc distance for the region, as well as be affected by the
tracer used for calculations.
High values of $\sigma_\textsc{nt}$
and Mach number have been determined
to be a result of large-scale accretion flows resulting in internal
turbulence, and when filament networks (with disparate velocity centroids)
are observed with low-resolution beams, the resulting measurements of
linewidth are expected to be supersonic \citep{Hacar_linewidth_2016AA}.

The central part of the Sh2-138 region has been mapped by JCMT \co, \tco, and
\cetno\, molecular lines. These molecular lines trace the warm and dense gas
-- in the temperature range 10-50\,K and density $\sim$\,10$^{4-5}$\,cm$^{-3}$ --
enveloping the cores where star formation is taking place
\citep{Davis_JLSGould_Taurus_2010}.
Three main filaments -- labelled W-f, SW-f, and SE-f --
can be traced on the \tco\, channel map (Figure \ref{fig_chmap_13CO32}).
While one end of SW-f and SE-f filaments merges into the hub, the morphology
at other end of these two filamentary structures shows other possible
filaments branching from them.
For example, in Figure \ref{fig_13co_spire250}, SW-f (``c8-c9-c10'') filament
seems to be branching into ``c10-c11-c12'' and ``c10-c13''; and SE-f seems to
be branching into ``c14-c15'' and ``c14-c16''. It is possible that they
represent secondary filaments which merge into primary filament (SW-f and SE-f
here), which then merges into the hub. Such structure has been seen for regions
like Monoceros R2 \citep{TrevinoMorales_MonR2_2019, Nanda_MonR2_2022AA}.
Furthermore, at the end of SW-f in Figure \ref{fig_13co_pv}(right), a clump
is seen at an offset of $\sim$\,0.052\,degrees (marked with a dashed vertical
line).
This clump corresponds to $\sim$\,c10/c11 in Figure \ref{fig_13co_spire250},
which seems to be accreting matter from further along the
filament length, given the small gradient which is seen in the 0.052-0.06\,degree
offset range in Figure \ref{fig_13co_pv}(right).
Along the length of the W-f and SW-f filaments marked in the
p-v diagrams Figures \ref{fig_13co_pv}(left) and \ref{fig_13co_pv}(right),
respectively, clumping of gas can also be seen.
The p-v maps (Figure \ref{fig_13co_pv}) show a velocity gradient as one
approaches the central clump region, which
indicates gas being channeled from outer regions to the central region.
The relatively large velocity dispersion in the central part of the cloud
in all three isotopologues (Figure \ref{fig_m012_combined}) is
indicative of this.
The scenario that emerges is that of
longitudinal flow along filaments converging on a stellar cluster
\citep{Peretto_SDC13_2014, TrevinoMorales_MonR2_2019, Nanda_2020}.

\begin{figure}
\includegraphics[width=\linewidth]{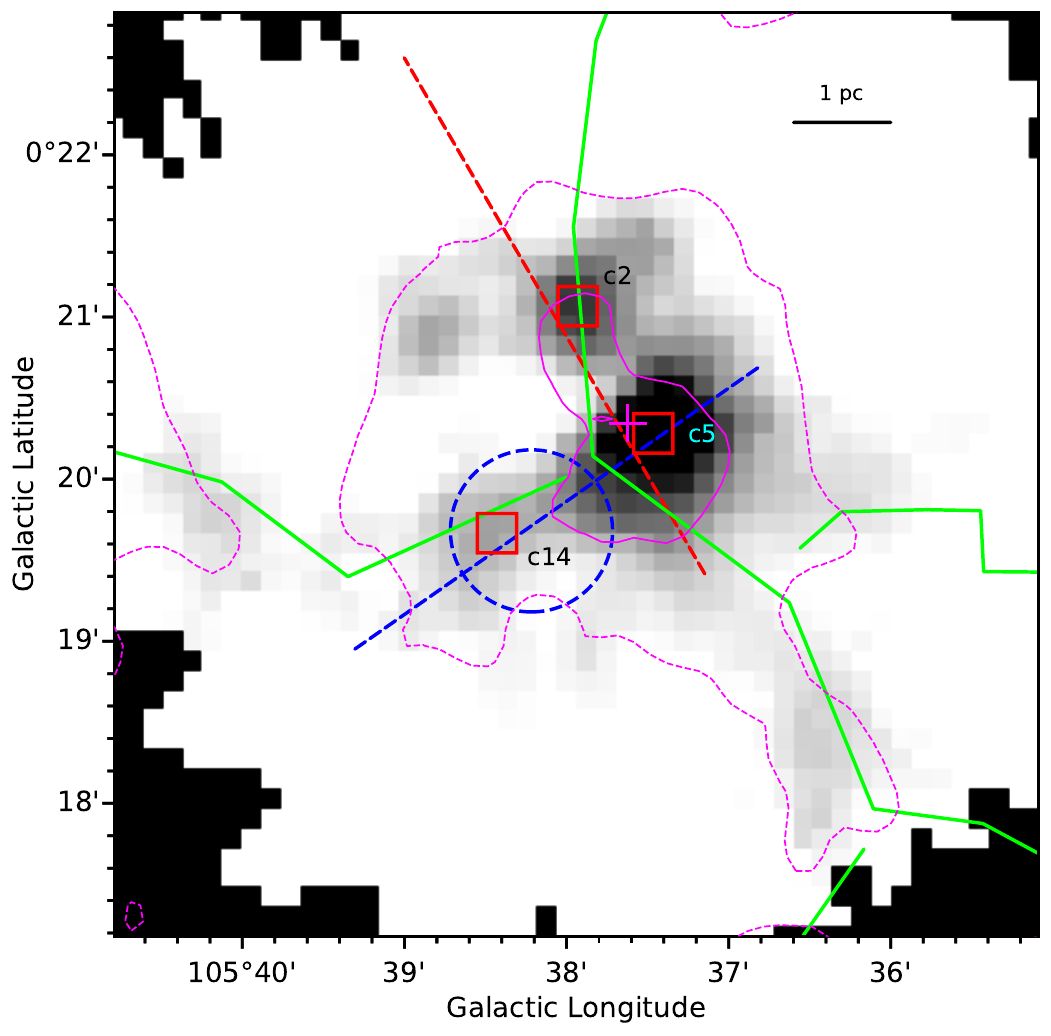}
\caption{JCMT \tco\, integrated intensity map.
Magenta dashed and solid contours mark N(H$_2$) = 33$\times$10$^{20}$\,cm$^{-2}$
and 10$^{22}$\,cm$^{-2}$, respectively, from the Vialactea maps. Dashed blue and
red lines mark the major axes of blue and red outflow contours, respectively,
from \citet{Qin_wrtS138_2008}.
Magenta plus symbol shows the location of IRAS\,22308+5812.
Green lines are filament skeletons from Figure \ref{fig_350getsf_central}.
The clumps c2, c5, and c14
from Figure \ref{fig_13co_spire250} have been
marked in by red boxes and labelled.}
\label{fig_jcmt13CO_combined}
\end{figure}

Figure \ref{fig_jcmt13CO_combined} shows the JCMT \tco\, integrated intensity
map with overlaid column density contours, outflow axes, and the \emph{getsf}
filamentary skeletons from Figure \ref{fig_350getsf_central} within this
field of view. The peak of the blueshifted outflow emission has been indicated
by a dashed blue circle and it coincides with the peak c14 from
Figure \ref{fig_13co_spire250}.
The trapezium-like cluster from \citet{Deharveng_S138_1999}
\citep[including a Herbig Be star;][]{Tapas_S138_2015},
is coincident with the IRAS source marked here, which being far away from the
blue lobe peak is unlikely to be the driving source of the blue outflow lobe.
The cluster of young stellar objects (in which some are of intermediate mass)
identified by \citet{Tapas_S138_2015} which lie within the area of the blue
lobe peak could be partially driving the outflow, or alternatively,
c14 could be a source in the initial stages of star formation. As such, this
source could be a suitable candidate as a subject of detailed investigation.
Comparing Figures \ref{fig_m012_combined} and \ref{fig_jcmt13CO_combined},
one can see that the northern (eastern) part of the  N-S (E-W) vector
in Figure \ref{fig_m012_combined}(e) is almost coincident with the
red (blue) lobe axis in Figure \ref{fig_jcmt13CO_combined}.
Thus the complex structure in the p-v diagram seen in
Figure \ref{fig_13co_pv}(right)
($\sim$ offset 0.00 to 0.02\,deg, corresponding to northern part of
the N-S vector of Figure \ref{fig_m012_combined}(e))
could be a combination of the turbulence injected along the directions
by the outflow and/or phenomena such as Hubble flow
\citep{RidgeMoore_AA_2001,Arce_HH300_ApJ_2001},
mass entrainment, and so on \citep{LadaFich_NGC2264G_ApJ_1996,Arce_PVCep_ApJ_2002}.
\citet{Zinchenko_S255IR_ApJ_2020} in their study of S255IR high-mass
star forming region have observed that walls around outflow cavities could
appear as filaments in projection. Given the arrangement of \emph{getsf}
filaments and outflow axes for the Sh2-138 region
(in Figure \ref{fig_jcmt13CO_combined}),
such a situation is also a possibility here and merits further investigation
of the region in different molecular species.
It is also worth noting that the central hub region (magenta solid contour
in Figure \ref{fig_jcmt13CO_combined}) seems to have two centers (c2 and c5),
also seen in other regions such as NGC\,2264 \citep{Nanda_2020} and
G31.41+0.31 \citep{Beltran_G314_2022AA}.

Looking at the combined large-scale field-of-view (Section \ref{section_largescaleview})
and the central portion (Section \ref{section_kinematics_JCMTobs})
from the context of star-formation frameworks, one finds that there could be
applicability of scenarios such as the
Global Hierarchical Collapse \citep[GHC][]{VazquezSemadeni_GHC_MNRAS_2019},
the ``Conveyor Belt'' model \citep{Longmore_ConveyorBelt_2014, Krumholz_ConveyorBelt_MNRAS_2020},
and the HFS model \citep{Nanda_2020} based on the longitudinal flows.
There also appears to be isolated star-formation all along the filament F1
in Figure \ref{fig_ratio_map}, and though the isolated 1.1\,mm emission
clump seems to be associated with a hub (based on the column density), it does
not come across as lying at any junction of filaments. A caveat, however, could
be that merging filaments at this clump are directed nearly orthogonal to
the plane of the sky in the line of sight.
Finally, we note that there are studies in literature which
-- based on molecular line spectra --
have suggested the existence of multiple clouds in various regions and
subsequent ``cloud-cloud-collision'', wherein collision between molecular
clouds leads to a shock-compressed layer with density enhancement,
due to which filament formation as well as high-mass star formation can occur
\citep{Scoville_CCC_1986,Habe_Ohta_CCC_1992,Tan_CCC_2000,Anathpindika_CCC_2010,
Inoue_Fukui_CCC_2013,Takahira_CCC_2014,Fukui_2021_CCC}.
Though multiple peaks in the CO spectra are seen for this region too
(Section \ref{section_largescaleview}), such a scenario however,
is tough to conclude, and would need observations
in molecules which trace the dense gas of shock-compressed layers,
such as NH$_3$ and HCN \citep{Priestly_Whitworth_CCC_2021} to justify.


\section{Summary and Conclusions}
\label{section_summaryconclusions}

In this paper we have carried out a molecular line study of the
``hub-filament'' system in the Sh2-138 region.
The primary data utilised was CO(1-0) transition from CGPS for the wider
$\sim$\,50\arcmin$\times$50\arcmin\, region,
and \co, \tco, and \cetno\, transitions from the JCMT archive for
the central $\sim$\,5\arcmin$\times$5\arcmin\, region.
The main conclusions are as follows :
\begin{enumerate}

\item
CGPS CO(1-0) integrated intensity emission of the region shows it
to be a HFS of $\sim$\,a few 10s of pc in scale. The central
region shows a spectrum with two velocity components.
The axes of outflows in the region from literature were found to
be aligned along the filaments detected via the \emph{getsf} tool.

\item
In the large-scale field-of-view, one of the filaments (labelled ``F1'' in
our analysis) appears to be a site of active star formation.
It is associated with diffused ionised emission at a junction with another
filament (labelled ``F2'') and was found to be hosting a 1.1\,mm emission clump
of mass $\sim$\,1606\,M$_\sun$.

\item
Analysis of the central $\sim$\,5\arcmin$\times$5\arcmin\, area in \tco\,
emission found three filamentary structures
-- labelled W-f, SW-f, and SE-f -- above a 5$\sigma$ detection threshold.
The observed line mass (M$_{\textup{line,obs}}$) was calculated to be
$\sim$\,50, 32, and 33.5\,M$_{\sun}$\,pc$^{-1}$ for SE-f, W-f, and SW-f,
respectively.

\item
The clump labelled c14, detected in \emph{Herschel} 250\,\micron\, emission
as well as \tco\, integrated intensity emission (at $>$\,5$\sigma$) was
found to coincide with the peak emission region of the blue outflow lobe,
and merits future investigation as a protostellar candidate.

\item
Position-velocity slices (east-west and north-south slice) across the
filaments revealed velocity gradients which point towards longitudinal
flow along the filaments converging onto the central dense clump.

\item
A gaussian model fitting of the spectra at different locations showed a
dominance of non-thermal motion -- with a large non-thermal dispersion
and a small value of the ratio of thermal to non-thermal pressure
($\sim$\,0.01--0.1).
Mach number ($\gtrsim$3) analysis indicates the
presence of large supersonic motions within the clumps.

\end{enumerate}


\section*{Acknowledgements}

We thank the anonymous referee for a critical reading of the manuscript
and for the suggestions for the improvement of this paper.
DKO acknowledges the support of the Department of Atomic Energy, Government
of India, under project Identification No. RTI 4002.
IIZ acknowledges support from IAP RAS program 0030-2021-0005.
The research work at Physical Research Laboratory is funded by the Department
of Space, Government of India.
TB acknowledges the support from S. N. Bose National Centre for Basic Sciences
under the Department of Science and Technology (DST), Government of India.
The Canadian Galactic Plane Survey (CGPS) is a Canadian project with
international partners. The Dominion Radio Astrophysical Observatory
is operated as a national facility by the National Research Council
of Canada. The Five College Radio Astronomy Observatory CO Survey of
the Outer Galaxy was supported by NSF grant AST 94-20159. The CGPS is
supported by a grant from the Natural Sciences and Engineering
Research Council of Canada.
The James Clerk Maxwell Telescope has historically been operated by the
Joint Astronomy Centre on behalf of the Science and Technology Facilities
Council of the United Kingdom, the National Research Council of Canada and
the Netherlands Organisation for Scientific Research.
This research has made use of the NASA/IPAC Infrared Science Archive,
which is funded by the National Aeronautics and Space Administration and
operated by the California Institute of Technology.

\facilities{JCMT, Herschel, IRSA}

\bibliographystyle{aasjournal}
\bibliography{bibliography.bib}
\end{document}